\documentclass[manuscript]{aastex}

\begin{document}

%% LaTeX will automatically break titles if they run longer than
%% one line. However, you may use \\ to force a line break if
%% you desire.

\title{Decreased Frequency of Strong Bars in S0 Galaxies: Evidence for Secular Evolution?}

%% Use \author, \affil, and the \and command to format
%% author and affiliation information.
%% Note that \email has replaced the old \authoremail command
%% from AASTeX v4.0. You can use \email to mark an email address
%% anywhere in the paper, not just in the front matter.
%% As in the title, you can use \\ to force line breaks.

\author{R. Buta\altaffilmark{1}, E. Laurikainen\altaffilmark{2,3}, H. Salo\altaffilmark{2}, and J. H. Knapen\altaffilmark{4,5}}
\altaffiltext{1}{Department of Physics and Astronomy, University of Alabama, Box 870324, Tuscaloosa, AL 35487}
\altaffiltext{2}{Department of Physics/Astronomy Division,
University of Oulu,  FIN-90014 Finland}
\altaffiltext{3}{Finnish Centre for Astronomy with ESO (FINCA), University of Turku}
\altaffiltext{4}{Instituto de Astrof\'i sica de Canarias, E-38200 La Laguna, Tenerife,
Spain}
\altaffiltext{5}{Departamento de Astrof\'i sica, Universidad de La Laguna, E-38205 La
Laguna, Tenerife, Spain}

%\email{aastex-help@aas.org}

%% Notice that each of these authors has alternate affiliations, which
%% are identified by the \altaffilmark after each name.  Specify alternate
%% affiliation information with \altaffiltext, with one command per each
%% affiliation.

\begin{abstract}
Using data from the Near-Infrared S0 Survey (NIRS0S) of nearby,
early-type galaxies, we examine the distribution of bar strengths in S0
galaxies as compared to S0/a and Sa galaxies, and as compared to
previously published bar strength data for Ohio State University Bright
Spiral Galaxy Survey (OSUBSGS) spiral galaxies. Bar strengths based on the
gravitational torque method are derived from 2.2$\mu$m $K_s$-band
images for a statistical sample of 138 (98 S0, 40 S0/a,Sa) galaxies
having a mean total blue magnitude $<B_T>$ $\leq$12.5 and generally
inclined less than 65$^o$. We find that S0 galaxies have weaker bars on
average than spiral galaxies in general, even compared to their closest
spiral counterparts, S0/a and Sa galaxies. The differences are
significant and cannot be due entirely to uncertainties in the assumed
vertical scale-heights or in the assumption of constant mass-to-light
ratios. Part of the difference is likely due simply to the dilution of
the bar torques by the higher mass bulges seen in S0s. If spiral
galaxies accrete external gas, as advocated by Bournaud \& Combes, then
the fewer strong bars found among S0s imply a lack of gas accretion
according to this theory.  If S0s are stripped former spirals, or else
are evolved from former spirals due to internal secular dynamical
processes which deplete the gas as well as grow the bulges, then the
weaker bars and the prevalence of lenses in S0 galaxies could further
indicate that bar evolution continues to proceed during and even after
gas depletion.

\end{abstract}

%% Keywords should appear after the \end{abstract} command. The uncommented
%% example has been keyed in ApJ style. See the instructions to authors
%% for the journal to which you are submitting your paper to determine
%% what keyword punctuation is appropriate.


%% From the front matter, we move on to the body of the paper.
%% In the first two sections, notice the use of the natbib \citep
%% and \citet commands to identify citations.  The citations are
%% tied to the reference list via symbolic KEYs. The KEY corresponds
%% to the KEY in the \bibitem in the reference list below. We have
%% chosen the first three characters of the first author's name plus
%% the last two numeral of the year of publication as our KEY for
%% each reference.

\section{Introduction}

Bars are the most important type of perturbation found in common among
spiral and S0 galaxies. In spiral galaxies where interstellar gas is
plentiful, a bar can be a major engine of gas-dominated secular
evolution, leading to radial gas flows, circumnuclear starbursts, and
possibly pseudobulges made of disk material (Kormendy \& Kennicutt
2004=KK04). In S0 galaxies, secular evolution is a different issue,
because such systems generally have far less interstellar gas than
spirals, and thus one must turn to possible interactions between the
stellar components, to external interactions, or to the possible
relationship between S0s and spirals, for evidence of secular
evolution.

Of particular interest is how the properties of bars in S0 galaxies
might differ from those in spirals. Bars in early-type spiral galaxies
have for some time been reported to be stronger and longer
than those in later-type galaxies (Elmegreen \& Elmegreen 1985).
However, such an assessment is usually based on contrast: the relative
$m$=2 Fourier intensity amplitude, $A_2$, is stronger for early-type
barred galaxies (Laurikainen et al. 2007), but when the {\it forcing}
due to a bar is taken into account, the bars of early-type spirals
actually come out weaker than those in later-type spirals (Buta,
Laurikainen, \& Salo 2004; Laurikainen, Salo, \& Buta 2004). This is
because bar strength is a relative parameter that depends on the radial
forcing due to the axisymmetric background (Combes \& Sanders 1981).
The axisymmetric contribution to the potential of early-type galaxies
tends to be strong due to the presence of more massive bulges
(Laurikainen et al. 2004, 2007). 

Until recently, there has been no detailed study capable of reliably
assessing how the distribution of bar strengths in S0 galaxies might
differ from those in both early and late-type spirals.  Our goal in
this paper is to examine this question using data from the
Near-Infrared S0 Survey (NIRS0S, Laurikainen et al 2005; Buta et al.
2006), a statistically well-defined $K_s$-band imaging survey of
183 S0 to Sa galaxies selected from the Third Reference Catalogue of
Bright Galaxies (RC3, de Vaucouleurs et al. 1991). The NIRS0S was
carried out from 2004-2009 at several major observatories, and some
analysis of the data has been presented by Laurikainen et al.  (2005,
2006, 2007, 2009, 2010a), focussed mainly on bulge properties of
early-type galaxies as opposed to later types.  We wish to examine the
implications of the distribution of bar strengths in S0 galaxies as
compared to that for spirals, in order to (1) further explore the
possible relationship between the two classes of objects, and (2)
deduce the evolutionary history of bars in the absence of long-term gas
flow.

\section{Data and Sample}

The NIRS0S was designed to complement and overlap the {\it Ohio State
University Bright Spiral Galaxy Survey} (OSUBSGS, Eskridge et al. 2002), a
sample of 205 spirals over the type range S0/a to Sm, with a
comparable-sized sample of early-type galaxies in the type range S0$^-$
to Sa. Owing to the lesser abundance of S0 galaxies compared to spirals among
bright galaxies, the selection criteria of the NIRS0S could not be made
identical to those of the OSUBSGS. The latter used a magnitude limit of
$B_T$ $\leq$ 12.0, while the NIRS0S uses a limit of $<B_T>$ $\leq$
12.5, where $<B_T>$ is the weighted mean of the mostly
photoelectrically-determined $B_T$ magnitudes in RC3 and photographic
magnitudes ($m_B$) on the same scale listed in the same
catalogue\footnote{This definition for the sample slightly differs from
that used by Buta et al. (2006), who adopted only a photoelectric
$B_T$$\leq$12.5. The reason for using $<B_T>$ is that many southern
galaxies only have total $V$ photoelectric magnitudes in RC3, but all
have $m_B$, allowing a better definition of the sample.}.  The OSUBSGS
had no inclination limit, but the NIRS0S galaxies were restricted to
log$R_{25}$ $\leq$ 0.35 in order to minimize deprojection
uncertainties. This limit nevertheless failed to exclude some edge-on
S0s, and even some galaxies just below the limit were found to be too
inclined to reliably deproject. In addition, NIRS0S includes 19
galaxies classified as types E or E$^+$ in RC3, but which are instead
classified as S0s in the {\it Revised Shapley-Ames Catalogue} (RSA,
Sandage \& Tammann 1981). We also included NGC 6482, which satisfies
the NIRS0S magnitude and diameter restrictions, but which is classed as
E$^+$ in the RC3 and E2 in the RSA. Our analysis shows it to be an S0.
The final sample includes 183 early-type galaxies and is listed in
Table 1. All details connected with the NIRS0S, including the public
availability of the images, will be presented by Laurikainen et al.
(2010b).

Buta et al. (2006) showed that the typical NIRS0S galaxy has an
absolute blue magnitude $M_B^o$ of $-$20.0, comparable to the OSUBSGS
bar strength sample (Buta et al. 2004). Dwarfs are greatly
under-represented in both samples and do not form part of our analysis
(see, e. g., Marinova \& Jogee 2007).

\section{Estimation of Bar Strengths}

We measure bar strengths from the maximum tangential forcing relative
to the mean background radial forcing (Combes \& Sanders 1981) using a
polar grid method (Salo et al. 1999; Laurikainen \& Salo 2002, 2005). A
deprojected near-infrared image is converted to a gravitational
potential assuming a constant mass-to-light ratio, and from this
potential the maximum ratio of the tangential to mean radial force,
$Q_T(r)$, is derived as a function of radius $r$ (an example is shown
in Figure 4 of Buta et al. 2005). The maximum of this function is
called $Q_g$, which is equivalent to the maximum gravitational torque
per unit mass per unit square of the circular speed. We use the polar
grid approach (as opposed to the Cartesian method used by Quillen,
Frogel, \& Gonzalez 1994) because it is less sensitive to noise and
provides more stable values of $Q_T(r)$ versus $r$. For the whole
NIRS0S sample, these functions will be presented by Salo et al.
(2010).

Although color-dependent mass-to-light ratio corrections can be made
based on empirical formulae from Bell \& de Jong (2001), we have made no
such corrections for the NIRS0S sample, and in any case the impact of
such corrections on bar strength estimates is small for early-type
galaxies because color gradients in such galaxies are generally small
(e.g., de Vaucouleurs et al. 1991). A more important problem could be
the contribution of dark matter to the axisymmetric background
potential. We showed in Buta, Laurikainen, \& Salo (2004) that
corrections to $Q_g$ for dark matter, based on a ``universal rotation
curve" (Persic, Salucci, \& Stel 1996), are relatively small for the
high luminosity OSUBSGS spirals. In the case of S0s, the situation is
likely to be similar.  For example, Williams et al. (2009) analyzed
rotation data for 14 S0s as compared to 14 Sa, Sb spirals and found no
systematic difference in the dark matter contents.

In this study we compare bar strengths of S0s with those we obtained
previously for spirals (Buta et al. 2004, 2005; Laurikainen et al.
2004). In barred spirals, $Q_g$ includes contributions from both the
bar and the spiral.  Therefore, for OSUBSGS spirals we also use bar
strengths in which the spiral contribution is eliminated using the
Fourier-based method of Buta, Block, \& Knapen (2003), which leads to
individual estimates of the bar strength, $Q_b$, and the spiral
strength, $Q_s$. In S0 galaxies, spiral structure is absent, 
$Q_b=Q_g$, and the two parameters can be used interchangeably. This
is also not a bad approximation even for spirals. For example, Figure 5
of Buta 2004 shows a separation for the SB0/a galaxy NGC 4596 where the
arms are too weak to make $Q_b$ different from $Q_g$. Also, Figure 26
of Buta et al. (2009) shows that even for stronger spirals, the
difference between $Q_g$ and $Q_b$ is generally small. $Q_g$ and $Q_b$
values for spirals in the OSUBSGS are provided by Laurikainen et al.
(2004) and Buta et al. (2005), respectively.

An important parameter in bar strength studies is the vertical scale
height, $h_z$, a quantity which can only be reliably measured in
edge-on galaxies, and which impacts the forcing in the plane used for
estimating bar strengths. In our previous analyses of OSUBSGS spiral
galaxies, we assumed an exponential vertical density distribution and
derived $h_z$ as a type-dependent fraction of the radial scale length
$h_R$, based on a study by de Grijs (1998). For types Sa and earlier,
Sab to Sbc, and Sc and later we used $h_z=h_R/4$, $h_R/5$, and $h_R/9$,
respectively. The de Grijs analysis includes very few early-type
galaxies, and for them we have used $h_z=h_R/4$ in order to make a
comparison with spirals. Values as low as $h_R/2$ or $h_R/3$ are also
possible for S0s from de Grijs's analysis.

The vertical scale height is the largest source of uncertainty in the
gravitational torque method (see Laurikainen \& Salo 2002). Because of
this, we also consider a different approach (Speltincx et al. 2008):
instead of scaling $h_z$ from the radial scale length, we scale it from
the $\mu_{K_s}$ = 20.0 mag arcsec$^{-2}$ isophotal radius, $r_{K20}$,
based on {\it Two-Micron All-Sky Survey} (2MASS; Skrutskie et al. 1997)
surface photometry provided on the NASA/IPAC Extragalactic Database
(NED) \footnote{The NASA/IPAC Extragalactic Database (NED) is operated
by the Jet Propulsion Laboratory, California Institute of Technology,
under contract with the National Aeronautics and Space Administration.}
website. We use scalings $h_z/r_{K20}$ = 0.05, 0.10, and 0.20, and
examine the full impact of $h_z$ assumptions on the distribution of
early-type galaxy bar strengths.

\section{Analysis}

\subsection{Sample Rejections}

Of the 183 NIRS0S galaxies listed in Table 1, 174\footnote{This number
excludes 31 additional S0-Sa galaxies that are outside the formal
definition of the sample.} were observed in our program (the objects
not observed are indicated in Table 1.) Of those observed, 35 had to be
rejected from our statistical analysis of bar strengths in early-type
galaxies either because they were too inclined in spite of what their
catalogued isophotal axis ratio implied, were interacting or peculiar,
or were members of close pairs where the two components were difficult
to separate. Nearby dwarfs such as NGC 205 and NGC 5206 were also
rejected. Of these, only NGC 5206 was actually observed. Also,
galaxies that, in the decomposition analysis of Laurikainen et al.
(2010a), could be fitted either by a single S\'ersic function, or a
combination of a S\'ersic function and some central component, were
rejected. Some of these are likely to be true elliptical galaxies,
although at least two of these cases (NGC 439 and 3706) are probably
not ellipticals. Table 1 identifies all the rejected galaxies and
gives the basis for the rejection. Our final sample for the bar
strength analysis consists of 138 galaxies.

\subsection{Distribution of Bar Strengths}

Our final statistical sample of 138 NIRS0S galaxies is comparable in
size to the 147 OSUBSGS galaxies used for bar and spiral strength
studies by Buta et al. (2005). Of the 138 galaxies, 98 are
classified as E-S0$^+$ in RC3 \footnote{In conflicting cases where a
galaxy was classified as type E in RC3 but S0 in the RSA, the RSA type
is adopted.}, while 40 are types S0/a and Sa. The latter types include
18 in common with the OSUBSGS (24 for the whole sample of 183 galaxies).

\begin{figure}
\figurenum{1}
\plotone{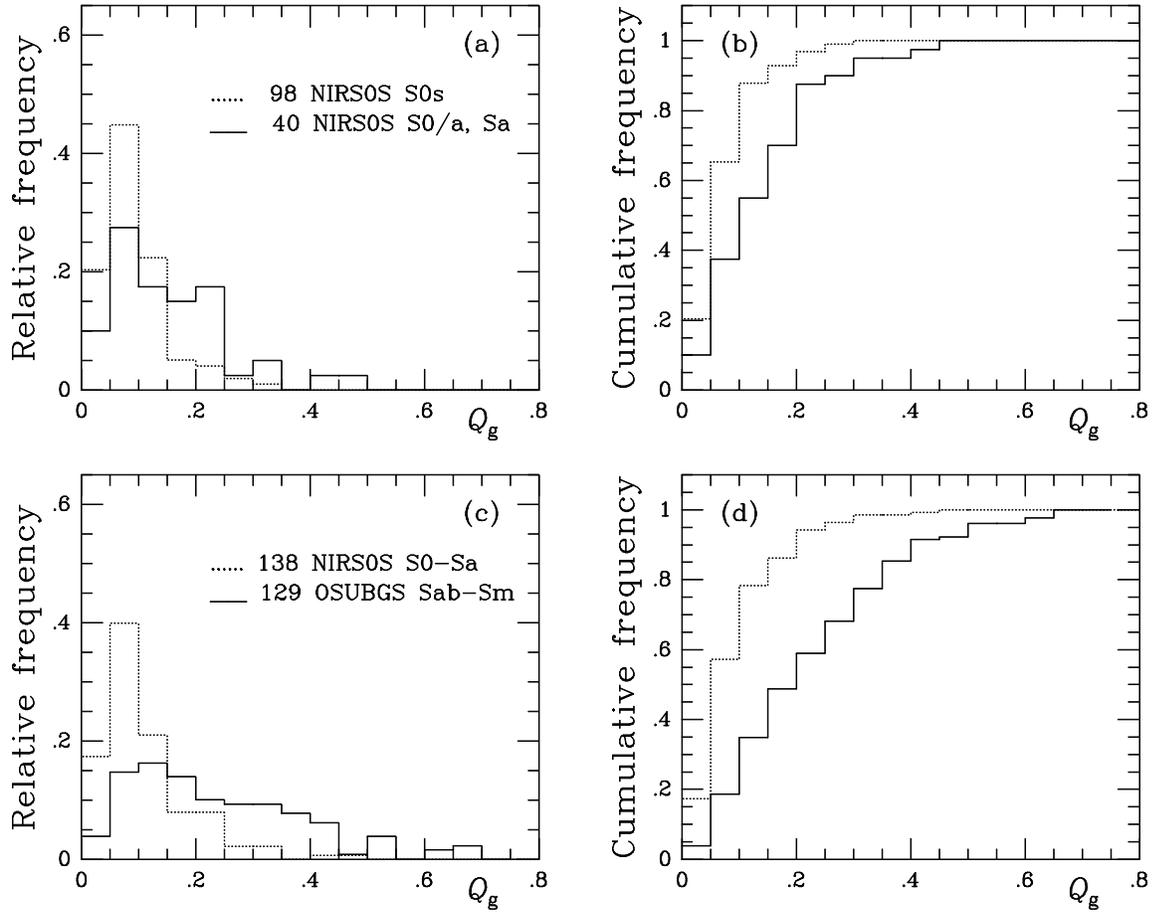}
\caption{(a,b) Histograms of the distribution of bar strengths in S0s
as compared early-type spirals (S0/a,Sa); (c,d) the same for the
early-type galaxy NIRS0S sample and a subset of OSUBSGS spirals.}
\label{bar1}
\end{figure}

Figure~\ref{bar1}a,b shows histograms (both differential and
cumulative) of the $Q_g$ relative frequency distributions of the S0 and
S0/a,Sa subsets of NIRS0S, and already it is evident that low $Q_g$
values are more abundant for S0s than for early-type spirals. For 98
S0s, $<Q_g>$ =0.09$\pm$0.06 (stand.  dev.), while for 40 S0/a and Sa
galaxies, $<Q_g>$ = 0.16$\pm$0.11 (stand. dev.). A Kolmogorov-Smirnov
test yields a $D$ parameter (the maximum difference between the
normalized cumulative distributions) of 0.34 for these subsamples, and
the corresponding probability $P$=0.002 indicates that the null
hypothesis that the two distributions come from the same parent
population is rejected at a high significance level. When the total
sample of 138 S0-Sa galaxies (Figure~\ref{bar1}c,d) is compared with
the distribution of $Q_g$ values for 129 OSUBSGS galaxies (Buta
et al. 2004, 2005), it is evident that S0s lack the extended ``tail" of
high $Q_g$ ($Q_b$) values noted by Block et al. (2002) for OSUBSGS
spirals. The Kolmogorov-Smirnov $D$ parameter is 0.51 for these
cumulative distributions, $P$$<$10$^{-3}$, and the null hypothesis is
again rejected. The OSUBSGS subset is restricted to the type range Sab
to Sm in order to exclude the 18 OSUBSGS S0/a and Sa galaxies in common
with the NIRS0S sample.

\begin{figure}
\figurenum{2}
\plotone{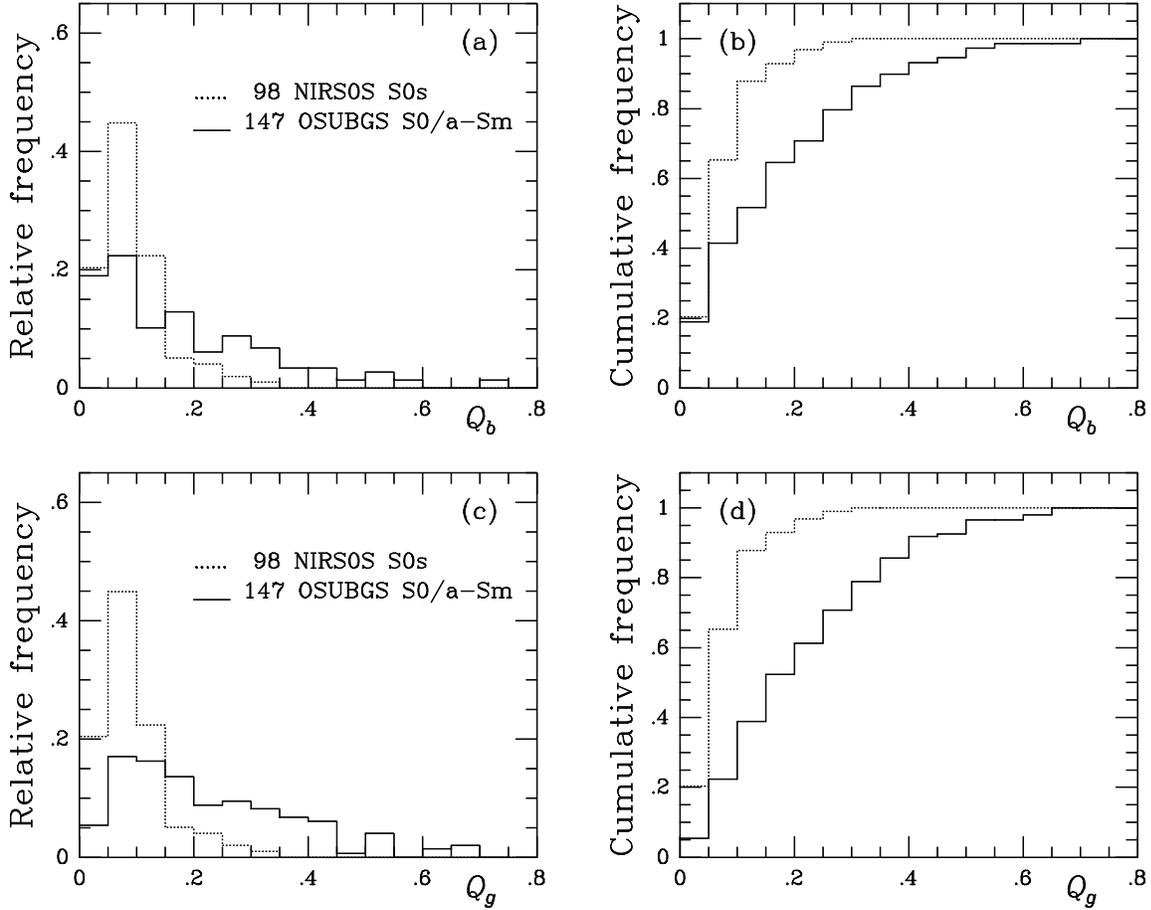}
\caption{Histograms of the distribution of bar strengths in S0s
as compared to OSUBSGS spirals, a sample dominated by intermediate
to late-type spirals. Note that the distinction between $Q_b$
and $Q_g$ is relevant only to the OSUBSGS sample. For the S0s,
we have set $Q_b$ = $Q_g$.}
\label{bar2}
\end{figure}

Figure~\ref{bar2} compares only the NIRS0S S0s with both $Q_g$ and
$Q_b$ distributions for 147 OSUBSGS S0/a-Sm spirals. The $Q_b$ parameter is
based on bar-spiral separations as described by Buta et al. (2005). The
presence of spiral torques exaggerates the difference between S0s and
OSUBSGS spirals because $Q_g$ is affected by spiral arm torques, but
when bar strength $Q_b$ alone is used, the distributions are still
significantly different. For the comparison with $Q_b$, the
Kolmogorov-Smirnov $D$=0.37, while for the comparison with $Q_g$, the
Kolmogorov-Smirnov is stronger at $D$=0.55. The corresponding $P$ in
each case is less than 10$^{-3}$, and the null hypothesis is again
rejected. The comparison with the OSUBSGS $Q_b$ values is most fair
since $Q_g$ for S0s will generally be a bar or oval strength without
significant contributions from spiral arms.

The differences found, between S0s and OSUBSGS spirals on one hand, and
between NIRS0S S0s and S0/a,Sa galaxies on the other, were previously
highlighted in a preliminary analysis of S0 bar strengths by Buta et
al. (2008). In that analysis, based partly on Sloan Digital Sky Survey
(SDSS) $i$-band images, no S0 galaxy having $Q_g$ $>$ 0.25 was found.
Of the three galaxies that have $Q_g$ $>$ 0.25 in our current analysis,
two [NGC 4596 ($Q_g$ = 0.28) and 4608 ($Q_g$ = 0.26)] are reclassified
as S0/a galaxies by Buta et al. (2007), and the third (NGC 4984) is
classified as Sa in the RSA.

\begin{figure}
\figurenum{3}
\plotone{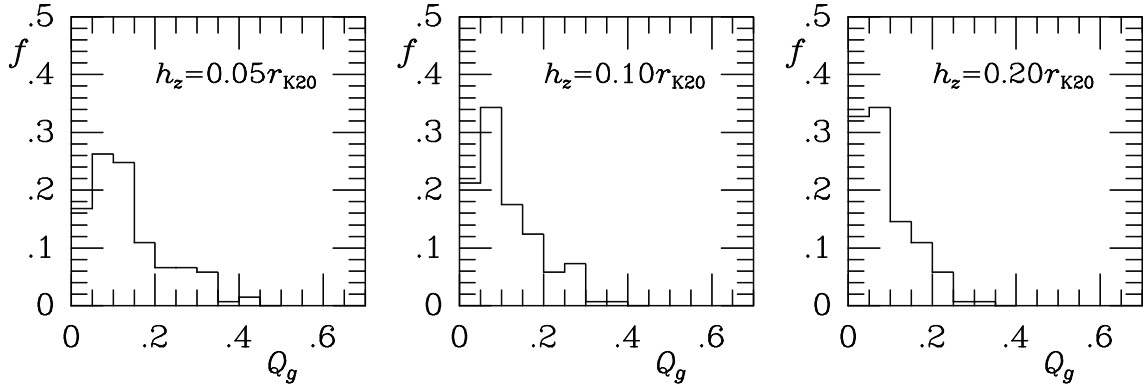}
\caption{Histograms of the distribution of bar strengths in 
137 NIRS0S galaxies for three different values of the assumed
vertical scaleheight.}
\label{bar3}
\end{figure}

Figure~\ref{bar3} shows histograms of the distribution of bar strengths
for the same galaxies as in Figures ~\ref{bar1} and ~\ref{bar2}, for
the alternative three values of the assumed scaleheight of the
exponential vertical density profile, in terms of fractions of the
isophotal radius $r_{K20}$. Of these, the one that is most like our
$h_z=h_r/4$ plots is for $h_z=0.10r_{K20}$, showing that our result is
not sensitive to how we define the scale height. The values
$h_z=0.05r_{K20}$ and $0.20r_{K20}$, are selected to logarithmically
bracket this value.  These highlight the significance of the assumed
thickness: as $h_z$ increases, the relative frequency of strong bars
decreases. If $h_z$ is as large as 0.20$r_{K20}$, there would be little
skewness or extended tail in the distribution of early-type galaxy bar
strengths.

This analysis can also give us an indication of what a factor of 2
uncertainty in the assumed average $h_z$ can do to the distributions.
A Kolmogorov-Smirnov comparison between the $h_z=0.05r_{K20}$ and
$h_z=0.10r_{K20}$ distributions gives a $D$-parameter of 0.16 with a
probability $P$=0.05 that the null hypothesis is rejected, more than a
factor of 100 poorer than the comparison shown in Figure~\ref{bar2}a,b
gives. Comparison between the $h_z=0.10r_{K20}$ and $h_z=0.20r_{K20}$
distributions gives a similar result. Thus, even a factor of 2
uncertainty in the assumed average $h_z$ would {\it not} rule out as
significantly that the samples come from the same parent population.

There can be little doubt that thickness effects contribute part of the
difference we find between spiral and S0 bar strengths.  Even so,
thickness can't fully explain the differences because even S0/a and Sa
galaxies have stronger bars on average than S0s. 

\section{Discussion}

Our study is the first to use the relative torque parameter $Q_g$
($Q_b$) for large well-defined samples to show the lower strengths of
bars in S0 galaxies as compared to spirals. This turns out to be in
good agreement with other ways of evaluating bar strength. For example,
Laurikainen et al. (2009) compared fractions of bars, ovals, and lenses
(in the near-IR) in galaxies of different types, and found that S0
galaxies have a smaller fraction of bars than S0/a galaxies or later
spirals. Aguerri et al. (2009) used isophotal ellipse fits 
(in the optical) to come to the same conclusion. Our goal in this
section is to examine several possible reasons for why the
distributions of bar strengths in spirals and S0s are so different. In
particular, we are interested in how the distributions connect to the
possibility that spirals are the progenitors of S0s, an idea which
originally came from comparison of the frequencies of S0s in
clusters and in the field (Dressler 1980).

\subsection{Do S0s have the bar strengths expected for systems not
accreting any gas?}

One way to interpret a distribution of bar strengths was discussed by
Bournaud \& Combes (2002), who used simulations to investigate the
possibility that a galaxy may have multiple bar episodes during a
Hubble time. If bar strength is a parameter that varies over time, then
the relative frequency of galaxies in each bin must tell us the
relative amount of time a galaxy spends in a given bar state
(non-barred, weakly-barred, or strongly-barred). The bar state is
something that could evolve due to gas accretion from external sources.
Bars first grow, perhaps by swing amplification (Combes 2000) or by
transferring angular momentum to a halo (Athanassoula 2003), and then
drive gas into the center, building up a central mass concentration.
This helps to heat the disk and weakens or destroys the bar. If there
is gas accretion, the disk can be cooled to the point of instability to
a new bar, as shown by Bournaud \& Combes (2002).  According to Block
et al. (2002), gas accretion produces an extended ``tail" to the
distribution of bar strengths, while lack of gas accretion leads to
a higher relative frequency of axisymmetric states than does gas
accretion. The distribution of bar strengths in S0 galaxies,
within the Bournaud \& Combes framework, seems to support the
idea that these galaxies have not accreted any external gas for a long
time, while spirals have accreted gas. However, it remains to be
determined in this framework what systematic environmental differences
result in differences in gas accretion efficiency. S0s may in general
have only a small amount of cool gas, but this does not explain why S0s
have not accreted more gas.

\subsection{Are weaker bars in S0s due mainly to more massive bulges
and thicker disks?}

S0 galaxies may not be directly comparable to the non-accreting models
of Bournaud \& Combes, which still have interstellar gas and spiral
structure. An alternative interpretation is that S0s have weaker bars
on average because they have more massive bulges than spirals
(Laurikainen et al. 2004, 2007). This dilutes the tangential forces,
making even a relatively strong-looking (high relative $m$=2 Fourier
amplitude or high maximum ellipticity) bar come out weak (see Figures 8
and 10 of Laurikainen, Salo, \& Buta 2004 for illustrations of this
dilution effect between early- and late-type spirals). As a check of
how much this effect plays a role, we analyzed the SB0 galaxy NGC 7155
from Buta et al. (2009), using similar techniques as for formal NIRS0S
galaxies. The bar strength is derived from a deprojected image obtained
by first subtracting a spherical bulge model, deprojecting the disk
component structures, and then adding back the bulge. The bar strength
obtained for this galaxy using such an image is 0.19. According to
Figure 1 of Laurikainen et al. (2007), an Sbc galaxy has about half the
$K_s$-band bulge-to-total luminosity ratio as a typical S0. In order to
simulate this, we add back to NGC 7155's disk component the same bulge
model scaled down by a factor.of 2. This yielded a bar strength of
0.26, 37\% larger than for the actual bulge model. Since the strongest
RC3 S0 bars have $Q_g$ $\leq$ 0.3, this kind of effect would translate
to a spiral maximum equivalent $Q_g$ $\leq$ 0.4, while the actual
spiral $Q_g$ can be as high as 0.7 (e.g., Laurikainen et al. 2004; Buta
et al. 2009).

De Grijs (1998) showed that on average, S0s have thicker disks than
late-type spirals. For a given mass, this implies less tangential
forcing in the plane.  The effect of the disk thickness on $Q_g$ was
estimated by Laurikainen \& Salo (2002): calculating two extreme cases
for the SBab galaxy NGC 1433, $h_z$=$h_R/2.5$ and 10, changes $Q_g$
from 0.4 to 0.5.

The combination of more massive bulges and thicker disks can therefore
account for some of the difference in average bar strength between S0s
and spirals. However, it is not likely to account for all of the
difference because other methods of estimating bar strength also
indicate that S0 bars are weaker on average than bars in spirals.
Aguerri et al. (2009) used the ellipticity-based $f_{bar}$ parameter
of Abraham \& Merrifield (2000) to show, for an SDSS volume-limited sample,
that S0 bars have a lower median $f_{bar}$ than early- or late-type
spirals.  Aguerri et al. (2009) tested the impact of the bulge on bar
ellipticities, and concluded it had no affect on their result that S0
bars are weaker than spiral bars.

\subsection{S0s as Transformed Spirals}

One of the main conclusions of Laurikainen et al. (2010a) is that the
specific photometric properties of S0 bulges and disks (e.g., S\'ersic
$n$, correlation of bulge and disk scale lengths) favor the idea that
S0s in general are evolved from spiral progenitors (see also Dressler
et al.  1997; Bekki et al. 2002; Shioya et al. 2004; van den Bergh
2009; and also Burstein et al. 2005 for a counter-argument). The
main problem with this interpretation is that S0 bulges are more
massive than spiral bulges, yet the present interstellar gas content of
spirals is insufficient to account for such masses (KK04). This could
simply mean that S0s had more gas-rich progenitors than present-day
spirals, or that another mechanism, unrelated to gas flow, could have
caused pre-existing disk stars to become part of a pseudobulge.  For
the latter, KK04 suggested that a bar buckling instability might be
relevant, although this mechanism mostly raises the stars away from the
disk plane, but does not necessarily lead to the radial migration of
the stars which is needed to build up the kind of central mass that
would significantly dilute $Q_g$. The concept of the bulge in the
buckling process is different from the kind of bulge we are concerned
with. A bulge created by buckling is actually part of the bar, as
discussed in many simulation studies (e.g., see review by Merrifield
1995).

If S0s are mainly spirals that have been stripped of (or used up, or
lost) all or most of their gas, then they should evolve as purely
stellar systems. Block et al. (2002) have noted: "In the absence of
gas, the dynamics of disks is different: pure stellar bars are very
robust, and can endure for one Hubble time, contrary to bars in
spirals. In S0s, bars are not destroyed, and no mechanism is needed to
explain bar reformation." This suggests that once the gas is gone, no
further evolution of the bar would be possible and the galaxy would
maintain the bar strength it had after the gas was removed. Thus,
whatever bar is left is not destroyed any further, and S0 galaxies
would be preserving a type of ``fossil" bar.

The more massive bulges in S0s nevertheless suggest, either that S0s
are not stripped spirals, or that secular evolution does occur in S0s,
including bar evolution. These bulges, which make S0 bars
systematically weaker compared to spirals, could be explained if there
existed an effective mechanism which allows the bulge to continue to
grow from disk material during, or even after, the gas depletion
process. One such mechanism, the potential-density phase shift (PDPS)
mechanism, was proposed by Zhang (1996, 1998, 1999).  According to the
PDPS mechanism, inward movement of the {\it stellar} component could
still occur as long as the bar wave mode has a small amount of skewness
(i.e., can be thought of as a very open spiral).  Just like a spiral,
bar skewness would introduce a phase shift between the density and
potential of the bar, causing stellar material (as well as gas) to
slowly move inward inside corotation and outward outside corotation,
with the result of slow secular mass increase in the bulge and a
spreading outward of the disk. A by-product of the radial mass
redistribution induced by the PDPS mechanism is the secular heating of
the disk stars in all three spatial dimensions, which also leads to the
growth of the bulge (Zhang 1999).

Zhang \& Buta (2007; their Figure 7 especially) show that slightly
skewed bars are indeed present among early-type barred galaxies, making
the PDPS mechanism a viable process to consider. The mechanism can
operate in any galaxy with a skewed pattern, and is purely
gravitational, so it affects both stars and gas. The mechanism is not
expected to be important if a bar is perfectly linear or a spiral is
very tightly wound (Zhang 2008). Note that the ability of the mechanism
to account for the more massive bulges in S0s does not depend on the
exact process that depletes the gas in a galaxy. Evolution from a later
type system (small $B/T$ ratio) to an earlier type system (larger $B/T$
ratio) follows naturally from the process.  If only star formation
depletes the gas, then the transition from spiral to S0 could be smooth
and the bulges of S0s could then simply be more advanced in an
evolutionary sense than those in spirals (Zhang 1999). If another
mechanism, such as ram pressure stripping, depletes the gas, the PDPS
mechanism could still operate, but the resulting S0 may not have as
large a $B/T$ mass ratio as it would have had if star formation
depleted the gas.

Studies of the role of a central mass concentration (CMC) in weakening
or destroying bars have suggested that a mere buildup of bulge mass
would be insufficient for total bar destruction. Shen \& Sellwood
(2004) used numerical simulations to show that diffuse CMCs are much
less effective at destroying bars than extremely dense CMCs.  In fact,
their simulations did not even use a bulge model, only a disk plus a
CMC. If an S0 did develop a massive enough (few percent disk mass) and
concentrated enough (supermassive black hole (SMBH)-like) CMC, then its
bar could be destroyed in a Hubble time. Athanassoula, Lambert, \&
Dehnen (2005; see also references therein) showed that a CMC with 10\%
of the mass of the disk could severely weaken a bar over time even in
the presence of a significant halo.

\subsection{Bars and Lenses}

Evidence that bars can be destroyed or weakened in S0s may come from
the existence of lenses. Kormendy (1979) showed that lenses (components
of galaxy structure having a shallow brightness gradient interior to a
sharp outer edge) are most abundant in SB0-SBa galaxies and much less
abundant in later-type barred galaxies, the latter having mostly inner
rings. Laurikainen et al. (2009) found lenses in 97\% of a subset of
127 NIRS0S galaxies; in that study also, S0/a galaxies included 82\%
with lenses.  Kormendy (1979) suggested that lenses are the products of
dissolved or dissolving bars, an idea supported by an observational
study of the lens in the S0$^+$ galaxy NGC 1553 (Kormendy 1984) and by
numerical simulations such as those of Bournaud \& Combes (2002).
Numerical simulations by Heller et al. (2007) have also shown that bars
can evolve to ``fat ovals" over time.  If this is indeed a viable
origin for lenses, then even if S0s were tied mainly to S0/a
progenitors, the lower relative frequency of lenses in S0/a types
compared to S0s suggests that bars do not ``freeze" after a galaxy is
stripped, but continue to evolve towards further dissolution.

Laurikainen et al. (2009) have suggested that the frequent presence of
lenses in S0s, of ansae in the bars of many S0s (Laurikainen et al.
2007; Martinez-Valpuesta et al. 2007), and of the double-peaked Fourier
profiles in early-type galaxies (Laurikainen et al. 2007), argues that
the bars of S0s are simply more evolved than those in spirals. If S0s
have spiral progenitors that did not necessarily have lenses at the
time the galaxies were stripped or cleaned of gas, then the formation
of some lenses had to have occurred {\it during} the S0 phase, meaning
bar evolution must have continued in the absence of much interstellar
gas.  A caveat connected with the lens issue is that not all lenses are
likely to be associated with dissolved bars. In galaxies with multiple
lenses, some could represent highly evolved former zones of active star
formation, such as inner, outer, and nuclear rings, in addition to
highly evolved bars.

\section{Conclusions}

We have highlighted the significant difference in the distribution of
bar strengths between S0 and spiral galaxies. Strong bars having $Q_b$
$>$ 0.25 are rare among S0s but relatively common among spirals.  Much
of the difference can be traced to the more massive bulges and thicker
disks on average that characterize S0s as compared to spirals, both of
which conspire to make S0 maximum relative gravitational bar torques low
compared to spirals. This cannot be the whole reason behind the
difference, however, because even when relative bar Fourier contrast,
bar ellipticity, and bar fraction are considered, bars are still less
prominent in S0s than in spirals. The suggestion is that if spirals
really are the progenitors of S0s, then there must exist a mechanism
that allows bar evolution to continue after gas depletion. One possible
mechanism, tied to the skewness of early-type galaxy bars, is a
potential-density phase shift that can evolve the purely stellar
distibution, allowing continued bulge-building that will help to
weaken any bar remaining after gas depletion. The earlier findings of
the high frequency of lenses in S0s fits to the scenario in which bar
evolution continues after depletion.

We thank X. Zhang for comments on an earlier version of this
manuscript. RB acknowledges the support of NSF Grant AST 05-07140 to
the University of Alabama.  HS and EL acknowledge the Academy of
Finland for support. JK acknowledges support by the Instituto de
Astrofisica de Canarias (312407).  This paper uses data obtained at
ES0/NTT (074.B0290(A), 077.A-0356(A), 081.B-0350(A)), as well at WHT,
TNG, NOT and KPNO.

\newpage
\centerline{REFERENCES}

\noindent
Abraham, R. G. \& Merrifield, M. R. 2000, \aj, 120, 2835

\noindent
Aguerri, J. A. L., M\'endez-Abreu, J., \& Corsini, E. M., 2009, \aap,
495, 491

\noindent
Athanassoula, E. 2003, \mnras, 341, 1179

\noindent
Athanassoula, E., Lambert, J. C., \& Dehnen, W. 2005, \mnras, 363, 496

\noindent
Bekki, K., Couch, W. J., \& Shioya, Y. 2002, \apj, 577, 651

\noindent
Bell, E. F. \& de Jong, R. S. 2001, \apj, 550, 212

\noindent
Bertin, G., Lin, C.C., Lowe, S.A., \& Thurstans, R.P. 1989,
ApJ, 338, 78

\noindent
Block, D. L., Bournaud, F., Combes, F., Puerari, I., \& Buta, R. 2002,
\aap, 394, L35

\noindent
Block, D. L., Freeman, K. C., Jarrett, T. H., Puerari, I., Worthey,
G., Combes, F., \& Groess, R. 2004, \aap, 425, L37

\noindent
Bournaud, F. \& Combes, F. 2002, \aap, 392, 83

\noindent
Burstein, D., Ho, L. C., Huchra, J. P., \& Macri, L. M. 2005, \apj, 621,
246

\noindent 
Buta, R. 2004, in Penetrating Bars Through Masks of Cosmic Dust,
D. L. Block, et al. eds., Dordrecht, Springer, p. 101

\noindent 
Buta, R. \& Block, D. L. 2001, \apj, 550, 243

\noindent
Buta, R., Block, D. L., and Knapen, J. H. 2003, \aj, 126, 1148

\noindent
Buta, R. J., Corwin, H. G., \& Odewahn, S. C. 2007, The de Vaucouleurs
Atlas of Galaxies, Cambridge: Cambridge U. Press

\noindent
Buta, R., Laurikainen, E., \& Salo, H. 2004, \aj, 127, 279

\noindent
Buta, R., Laurikainen, E., Salo, H., Block, D. L., \& Knapen, J. H.
2006, \aj, 132, 1859

\noindent
Buta, R., Laurikainen, E., Salo, H., Knapen, J. H., \& Block, D. L.
2008, in Formation and Evolution of Galaxy Bulges, Proceedings of the
International Astronomical Union, IAU Symposium, Vol. 245, p. 131

\noindent
Buta, R. J., Knapen, J. H., Elmegreen, B. G., Salo, H.,
Laurikainen, E., Elmegreen, D. M., Puerari, I., \& Block,
D. L. 2009, \aj, 137, 4487

\noindent Buta, R., Vasylyev, S., Salo, H., and Laurikainen, E. 2005, \aj, 130,
506

\noindent
Buta, R. \& Zhang, X. 2009, \apjs, 182, 559

\noindent
Combes, F. 2000, ASPC, 197, 15

\noindent
Combes, F. \& Sanders, R. H. 1981, \aap, 96, 164

\noindent
de Grijs, R. 1998, \mnras, 299, 595

\noindent
de Vaucouleurs, G., de Vaucouleurs, A., Corwin, H. G., Buta, R. J.,
Paturel, G., \& Fouque, P. 1991, Third Reference Catalog of Bright Galaxies (New
York: Springer) (RC3)

\noindent
Dressler, A. 1980, \apj, ApJ, 236, 351

\noindent
Dressler, A., Oemler, A., Couch, W. J., Smail, I., Ellise, R. S.,
Barger, A., Butcher, H., Poggianti, B. M., \& Sharples, R. M. 1997,
\apj, 490, 577

\noindent
Elmegreen, B. G. \& Elmegreen, D. M. 1985, \apj, 288, 438

\noindent
Eskridge, P. B., Frogel, J. A., Pogge, R. W., et al. 2002, ApJS, 143, 73

\noindent
Heller, C., Shlosman, I., \& Athanassoula, E. 2007, \apj, 671, 226

\noindent
Kormendy, J. 1979, \apj, 227, 714

\noindent
Kormendy, J. \& Kennicutt, R. 2004, ARAA, 42, 603 (KK04)

\noindent
Laurikainen, E. \& Salo, H. 2002, \mnras, 337, 1118

\noindent
Laurikainen, E., Salo, H., \& Buta, R. 2004, \apj, 607, 103

\noindent
Laurikainen, E., Salo, H., \& Buta, R. 2005, \mnras, 362, 1319

\noindent
Laurikainen, E., Salo, H., Buta, R., \& Vasylyev, S. 2004, \mnras, 355,
1251

\noindent
Laurikainen, E., Salo, H., Buta, R., Knapen, J., Speltincx, T.,
\& Block, D. L.  2006, \mnras, 132, 2634

\noindent
Laurikainen, E., Salo, H., Buta, R., \& Knapen, J. H. 2007,
\mnras, 381, 401

\noindent
Laurikainen, E., Salo, H., Buta, R., \& Knapen, J. H. 2009,
\apj, 392, L34

\noindent
Laurikainen, E., Salo, H., Buta, R., Knapen, J. H., \& Comeron, S. 2010a,
in press, \mnras (astro-ph 1002.4370)

\noindent
Laurikainen, E., Salo, H., Buta, R., \& Knapen, J. H. 2010b,
in preparation

\noindent
Marinova, I. \& Jogee, S. 2007, \apj, 659, 1176

\noindent
Martinez-Valpuesta, I. Knapen, J. H., \& Buta, R. 2007, \aj, 134, 1863

\noindent
Merrifield, M. R. 1995, ASP Conf. Ser. Vol. 91, p. 179

\noindent 
Persic, M., Salucci, P., \& Stel, F. 1996, \mnras, 281, 27

\noindent 
Quillen, A. C., Frogel, J. A., \& Gonzalez, R. 1994, \apj, 437, 162

\noindent 
Sandage, A. and Tammann, G. A. (1981), {\it A Revised
Shapley-Ames Catalog of Bright Galaxies}, Carnegie Institute of
Washington Publ. No. 635 (first
edition).

\noindent
Salo, H., et al. 2010, in preparation

\noindent 
Sanders, R. H. \& Tubbs, A. D. 1980, \apj, 235, 803

\noindent
Shen, J. \& Sellwood, J. A. 2004, \apj, 604, 614

\noindent
Shioya, Y., Bekki, K., \& Couch, W. J. 2004, \apj, 601, 654

\noindent
Skrutskie, M. et al. 1997, ASSL, 210, 25

\noindent
Speltincx, T., Laurikainen, E., \& Salo, H. 2008, \mnras, 383, 317

\noindent
van den Bergh, S. 2009, \apj, 702, 1502

\noindent
Williams, M. J., Bureau, M., \& Cappellari, M. 2009, \mnras, 400, 1665

\noindent
Zhang, X. 1996, ApJ, 457, 125

\noindent
Zhang, X. 1998, ApJ, 499, 93

\noindent
Zhang, X. 1999, ApJ, 518, 613

\noindent
Zhang, X. 2008, \pasp, 120, 121

\noindent
Zhang, X. \& Buta, R. 2007, \aj, 133, 2584

\newpage

\begin{deluxetable}{lrcclrcclrcc}
\tablenum{1}
\tablewidth{43pc}
\tablecaption{The NIRS0S Sample of 183 Galaxies\tablenotemark{a}}
\tablehead{
\colhead{Galaxy} &
\colhead{$T$} &
\colhead{$Q_g$} &
\colhead{$\sigma$} &
\colhead{Galaxy} &
\colhead{$T$} &
\colhead{$Q_g$} &
\colhead{$\sigma$} &
\colhead{Galaxy} &
\colhead{$T$} &
\colhead{$Q_g$} &
\colhead{$\sigma$} 
\\
\colhead{1} &
\colhead{2} &
\colhead{3} &
\colhead{4} &
\colhead{1} &
\colhead{2} &
\colhead{3} &
\colhead{4} &
\colhead{1} &
\colhead{2} &
\colhead{3} &
\colhead{4} 
}
\startdata
N0205\tablenotemark{b} & $-$5.0 & ..... & ..... & N3384                  & $-$3.0 & 0.055 & 0.019 & N4772                  &  1.0 & 0.061 & 0.001 \\
N0404\tablenotemark{b} & $-$3.0 & ..... & ..... & N3412                  & $-$2.0 & 0.081 & 0.013 & N4880                  & $-$1.0 & 0.115 & 0.016 \\
N0439\tablenotemark{h} & $-$3.3 & ..... & ..... & N3414\tablenotemark{c} & $-$2.0 & 0.084 & 0.042 & N4914                  & $-$4.0 & 0.064 & 0.004 \\
N0474                  & $-$2.0 & 0.054 & 0.016 & N3489                  & $-$1.0 & 0.068 & 0.008 & N4976                  & $-$5.0 & 0.045 & 0.003 \\
N0507                  & $-$2.0 & 0.052 & 0.008 & N3516                  & $-$2.0 & 0.066 & 0.009 & N4984                  & $-$1.0 & 0.340 & 0.006 \\
N0524                  & $-$1.0 & 0.019 & 0.004 & N3607                  & $-$2.0 & 0.088 & 0.000 & N5026                  &  $-$.2 & 0.222 & 0.062 \\
N0584                  & $-$5.0 & 0.098 & 0.025 & N3619                  & $-$1.0 & 0.024 & 0.006 & N5078\tablenotemark{i} &  1.0 & ..... & ..... \\
N0718                  &  1.0 & 0.121 & 0.009 & N3626                  & $-$1.0 & 0.106 & 0.000 & N5087                  & $-$3.0 & 0.171 & 0.000 \\
N0890                  & $-$3.0 & 0.088 & 0.001 & N3665                  & $-$2.0 & 0.078 & 0.013 & N5101                  &  0.0 & 0.237 & 0.016 \\
N0936                  & $-$1.0 & 0.204 & 0.045 & N3706\tablenotemark{h} & $-$3.0 & ..... & ..... & N5121                  &  1.0 & 0.025 & 0.003 \\
N1022                  &  1.0 & 0.145 & 0.009 & N3718\tablenotemark{g} &  1.0 & 0.315 & 0.059 & N5128\tablenotemark{b} & $-$2.0 & ..... & ..... \\
N1079                  &  0.0 & 0.228 & 0.039 & N3729                  &  1.0 & 0.216 & 0.003 & N5206\tablenotemark{j} & $-$2.5 & 0.150 & 0.010 \\
N1161                  & $-$2.0 & 0.127 & 0.009 & N3892                  & $-$1.0 & 0.190 & 0.008 & N5266                  & $-$3.0 & 0.074 & 0.001 \\
N1201                  & $-$2.0 & 0.101 & 0.017 & N3900                  & $-$1.0 & 0.081 & 0.021 & N5273                  & $-$2.0 & 0.036 & 0.012 \\
N1291                  &  0.0 & 0.102 & 0.009 & N3941                  & $-$2.0 & 0.113 & 0.010 & N5353\tablenotemark{d} & $-$2.0 & 0.246 & 0.002 \\
N1302                  &  0.0 & 0.092 & 0.006 & N3945                  & $-$1.0 & 0.094 & 0.009 & N5354\tablenotemark{d} & $-$2.0 & ..... & ..... \\
N1316                  & $-$2.0 & 0.076 & 0.002 & N3998                  & $-$2.0 & 0.037 & 0.008 & N5377                  &  1.0 & 0.244 & 0.069 \\
N1317                  &  1.0 & 0.082 & 0.002 & N4073\tablenotemark{h} & $-$3.8 & ..... & ..... & N5365                  & $-$3.0 & 0.105 & 0.009 \\
N1326                  & $-$1.0 & 0.139 & 0.009 & N4105\tablenotemark{d} & $-$5.0 & 0.092 & 0.002 & N5419                  & $-$4.7 & 0.052 & 0.015 \\
N1344                  & $-$5.0 & 0.081 & 0.011 & N4106\tablenotemark{d} & $-$1.0 & ..... & ..... & N5448\tablenotemark{i} &  1.0 & ..... & ..... \\
N1350                  &  1.8 & 0.210 & 0.050 & N4138                  & $-$1.0 & 0.047 & 0.002 & N5473                  & $-$3.0 & 0.082 & 0.007 \\
N1371                  &  1.0 & 0.115 & 0.015 & N4143                  & $-$2.0 & 0.075 & 0.006 & N5485                  & $-$2.0 & 0.043 & 0.002 \\
N1380                  & $-$2.0 & 0.158 & 0.003 & N4150                  & $-$2.0 & 0.053 & 0.003 & N5493\tablenotemark{c} & $-$2.0 & 0.291 & 0.005 \\
N1387                  & $-$3.0 & 0.067 & 0.003 & N4203                  & $-$3.0 & 0.040 & 0.002 & N5631                  & $-$2.0 & 0.051 & 0.005 \\
N1389                  & $-$3.3 & 0.080 & 0.018 & N4245                  &  0.0 & 0.188 & 0.015 & N5701                  &  0.0 & 0.217 & 0.001 \\
N1400                  & $-$3.0 & 0.034 & 0.001 & N4262                  & $-$3.0 & 0.071 & 0.024 & N5728                  &  1.0 & 0.343 & 0.003 \\
N1411                  & $-$3.0 & 0.022 & 0.006 & N4267                  & $-$3.0 & 0.045 & 0.003 & N5846\tablenotemark{h} & $-$5.0 & ..... & ..... \\
N1512                  &  1.0 & 0.158 & 0.006 & N4281\tablenotemark{h} & $-$1.0 & ..... & ..... & N5898                  & $-$5.0 & 0.023 & 0.001 \\
N1533                  & $-$3.0 & 0.105 & 0.008 & N4293\tablenotemark{i} &  0.0 & ..... & ..... & N6340                  &  0.0 & 0.029 & 0.001 \\
N1537                  & $-$2.5 & 0.146 & 0.063 & N4314                  &  1.0 & 0.445 & 0.010 & N6438\tablenotemark{f} & $-$2.0 & ..... & ..... \\
N1543                  & $-$2.0 & 0.125 & 0.017 & N4339                  & $-$5.0 & 0.030 & 0.005 & N6482                  & $-$5.0 & 0.064 & 0.002 \\
N1546\tablenotemark{b} & $-$1.3 & ..... & ..... & N4340                  & $-$1.0 & 0.224 & 0.032 & N6684                  & $-$2.0 & 0.109 & 0.001 \\
N1553                  & $-$2.0 & 0.102 & 0.000 & N4350\tablenotemark{c} & $-$2.0 & ..... & ..... & N6703                  & $-$2.5 & 0.022 & 0.001 \\
N1574                  & $-$2.7 & 0.065 & 0.006 & N4369                  &  1.0 & 0.491 & 0.005 & N6861\tablenotemark{i} & $-$3.0 & 0.155 & 0.016 \\
N1617                  &  1.0 & 0.100 & 0.028 & N4371                  & $-$1.0 & 0.246 & 0.002 & N6958                  & $-$3.8 & 0.023 & 0.006 \\
N1808\tablenotemark{b} &  1.0 & ..... & ..... & N4373\tablenotemark{d} & $-$2.9 & 1.057 & 0.037 & N7029                  & $-$5.0 & 0.083 & 0.000 \\
N1947\tablenotemark{b} & $-$3.0 & ..... & ..... & N4378                  &  1.0 & 0.053 & 0.000 & N7049                  & $-$2.0 & 0.084 & 0.000 \\
N2196                  &  1.0 & 0.065 & 0.023 & N4382\tablenotemark{k} & $-$1.0 & ..... & ..... & N7079                  & $-$2.0 & 0.084 & 0.008 \\
N2217                  & $-$1.0 & 0.110 & 0.002 & N4406\tablenotemark{l} & $-$5.0 & ..... & ..... & N7098                  &  1.0 & 0.160 & 0.004 \\
N2273                  &  0.5 & 0.196 & 0.008 & N4424\tablenotemark{i} &  1.0 & ..... & ..... & N7192                  & $-$4.3 & 0.010 & 0.004 \\
N2292\tablenotemark{d} & $-$2.0 & ..... & ..... & N4429\tablenotemark{i} & $-$1.0 & 0.283 & 0.002 & N7213                  &  1.0 & 0.010 & 0.000 \\
N2293\tablenotemark{d} & $-$1.0 & ..... & ..... & N4435\tablenotemark{c} & $-$2.0 & 0.130 & 0.001 & N7371                  &  0.0 & 0.075 & 0.005 \\
N2300                  & $-$2.0 & 0.056 & 0.012 & N4457                  &  0.0 & 0.108 & 0.008 & N7377                  & $-$1.0 & 0.052 & 0.007 \\
N2380                  & $-$1.7 & 0.023 & 0.005 & N4459                  & $-$1.0 & 0.027 & 0.003 & N7457                  & $-$3.0 & 0.050 & 0.022 \\
N2655                  &  0.0 & 0.089 & 0.004 & N4474\tablenotemark{c} & $-$2.0 & 0.094 & 0.013 & N7585                  & $-$1.0 & 0.070 & 0.010 \\
N2681                  &  0.0 & 0.058 & 0.001 & N4477                  & $-$2.0 & 0.100 & 0.025 & N7727                  &  1.0 & 0.075 & 0.015 \\
N2685\tablenotemark{g} & $-$1.0 & 0.249 & 0.003 & N4503                  & $-$3.0 & 0.061 & 0.015 & N7743                  & $-$1.0 & 0.109 & 0.019 \\
N2768\tablenotemark{i} & $-$5.0 & 0.072 & 0.008 & N4531                  & $-$0.5 & 0.119 & 0.001 & N7796                  & $-$3.8 & 0.053 & 0.001 \\
N2781                  & $-$1.0 & 0.065 & 0.006 & N4552                  & $-$5.0 & 0.137 & 0.039 & I1392                   & $-$3.0 & 0.120 & 0.021 \\
N2782                  &  1.0 & 0.109 & 0.067 & N4578                  & $-$2.0 & 0.063 & 0.000 & I4214                   &  1.5 & 0.158 & 0.008 \\
N2787                  & $-$1.0 & 0.172 & 0.044 & N4596                  & $-$1.0 & 0.281 & 0.047 & I4329                   & $-$3.0 & 0.031 & 0.007 \\
N2859                  & $-$1.0 & 0.102 & 0.004 & N4608                  & $-$2.0 & 0.263 & 0.010 & I4889                   & $-$5.0 & 0.121 & 0.024 \\
N2911                  & $-$2.0 & 0.079 & 0.002 & N4612                  & $-$2.0 & 0.086 & 0.000 & I4991                   & $-$2.0 & 0.063 & 0.004 \\
N2950                  & $-$2.0 & 0.103 & 0.069 & N4638\tablenotemark{c} & $-$3.0 & 0.251 & 0.020 & I5240                   &  1.0 & 0.192 & 0.016 \\
N3100                  & $-$2.0 & 0.064 & 0.031 & N4643                  &  0.0 & 0.332 & 0.005 & I5250 \tablenotemark{b} & $-$2.0 & ..... & ..... \\
N3166                  &  0.0 & 0.216 & 0.004 & N4649\tablenotemark{d} & $-$5.0 & 0.354 & 0.015 & I5250A\tablenotemark{b} & $-$2.0 & ..... & ..... \\
N3169                  &  1.0 & 0.082 & 0.009 & N4665                  &  0.0 & 0.257 & 0.016 & I5267                   &  0.0 & 0.026 & 0.001 \\
N3226\tablenotemark{d} & $-$5.0 & 0.098 & 0.001 & N4691\tablenotemark{e} &  0.0 & ..... & ..... & I5328                   & $-$5.0 & 0.063 & 0.003 \\
N3227\tablenotemark{d} &  1.0 & 0.185 & 0.013 & N4694                  & $-$2.0 & 0.141 & 0.007 & E137$-$010                  & $-$2.7 & 0.056 & 0.009 \\
N3245                  & $-$2.0 & 0.062 & 0.002 & N4696                  & $-$4.0 & 0.041 & 0.003 & E137$-$034\tablenotemark{b} &  0.0 & ..... & ..... \\
N3358                  &  0.0 & 0.088 & 0.008 & N4754                  & $-$3.0 & 0.199 & 0.028 & E208$-$021\tablenotemark{h} & $-$3.0 & ..... & ..... \\
\enddata
\noindent
\tablenotetext{a}{Col. 1: galaxy name; col. 2: RC3 numerical stage
index; col. 3: bar strength; col 4: error estimate for bar strength.
The sample is selected according to mean total blue magnitude
$<B_T>$ $\leq$ 12.5, logarithmic isophotal blue light axis ratio
$logR_{25} \leq$ 0.35, and RC3 numerical stage index $T$$<$2 (earlier
than Sab).}
\tablenotetext{b}{no observations obtained in program}
\tablenotetext{c}{edge-on galaxy}
\tablenotetext{d}{member of a close pair}
\tablenotetext{e}{no clear nucleus}
\tablenotetext{f}{strongly interacting}
\tablenotetext{g}{peculiar or otherwise disturbed}
\tablenotetext{h}{luminosity distribution well-fitted by a single Sersic function}
\tablenotetext{i}{inclination too high for reliable deprojection}
\tablenotetext{j}{dwarf}
\tablenotetext{k}{peculiar, nonbarred, edge-on, and warped?}
\tablenotetext{l}{no bar; flattened bulge}
\end{deluxetable}
\newpage

\end{document}